%% file: main.tex
\pdfoutput=1
\documentclass[11pt]{article}

\usepackage[margin=1in]{geometry}
\usepackage[T1]{fontenc}
\usepackage{lmodern}
\usepackage{microtype}
\usepackage{graphicx}
\usepackage{multirow}
\usepackage{amsmath,amssymb,amsfonts}
\usepackage{amsthm}
\usepackage{booktabs}
\usepackage{xcolor}
\usepackage{textcomp}
\usepackage{placeins}
\usepackage{rotating}
\usepackage[font=small,labelfont=bf]{caption}
\usepackage[numbers,sort&compress]{natbib}
\usepackage[colorlinks=true,linkcolor=blue,citecolor=blue,urlcolor=blue]{hyperref}
\theoremstyle{plain}
\newtheorem{proposition}{Proposition}
\input{numbers}

\hypersetup{pdftitle={Know Before You Fetch: When Does Confidence-Based Retrieval-Budget Allocation Pay Off?}, pdfauthor={Zhe Dong; Fang Qin; Manish Shah}}

\begin{document}

\title{Know Before You Fetch: When Does Confidence-Based Retrieval-Budget Allocation Pay Off?}

\author{%
Zhe Dong\\
University of Maine at Presque Isle\\
Presque Isle, ME, USA
\and
Fang Qin\\
Stanford University\\
Stanford, CA, USA
\and
Manish Shah\\
Independent Researcher\\
San Jose, CA, USA
}
\date{}

\maketitle

\begin{abstract}
Adaptive retrieval-augmented generation uses a signal from the reader, such as its confidence when answering without retrieval (closed-book), to decide whether and how much to retrieve. We compare such methods not with always retrieving many passages but with the best simple policy under the same cap on average passages per query: a fixed depth or a random mixture of fixed depths. Policies allocate zero, one or five passages and are scored on held-out questions. Seven settings of about 2,000 questions combine TriviaQA, Natural Questions and MS MARCO, Qwen3-8B and Qwen3-32B readers, and two retrieval sources: per-question passage pools and a 3.5-million-passage Wikipedia index. Allocation by confidence could gain in three ways. It rations a scarce budget: at a cap of half a passage per query it beats the best simple policy in all seven settings, by \gHalfMin{} to \gHalfMax{} in accuracy. It avoids harmful retrieval: at caps of one passage or more, substantial gains appear only on Wikipedia, where passages often overwrite known answers, most robustly for Qwen3-8B on TriviaQA (\gTqaWikiETwo{}, also under exact match and F1). It barely helps to choose depth: closed-book confidence ranks which queries a first passage helps but ranks the choice between one and five passages little better than chance, and with passage pools a fixed depth of two or three passages is as good or better. Once the probe's latency is charged, gains remain only where the gate skips retrieval for most queries; for the larger reader they also depend on the answer metric.
\end{abstract}


\section{Introduction}\label{sec:intro}

A retrieval-augmented question-answering system must decide, for every query, how much retrieved text to give its reader. Each passage adds input tokens and time. A passage can supply a fact the reader lacks, but it can also distract the reader from an answer it already knew \cite{cuconasu2024power,wang2024astute}. Adaptive retrieval methods respond by spending retrieval only where it seems useful. Most of them ask the reader first: they take its confidence in a closed-book answer, its internal states or the output of a trained classifier as a signal, and retrieve when the reader appears unsure \cite{jiang2023flare,asai2024selfrag,jeong2024adaptiverag,wang2023skr,wang2026targ,moskvoretskii2025adaptive}.

Such methods are usually judged against a system that retrieves a fixed and generous number of passages, often five or more, for every query. The comparison leaves out two cheaper alternatives that need no per-query decision. One is a smaller fixed depth: if one or two passages already give most of the accuracy of five, a system can simply read fewer passages for every query. The other is a random mixture of fixed depths, for example one passage for a randomly chosen half of the queries and none for the rest. A mixture can spend the same average number of passages as the adaptive method without looking at the query, so an adaptive method is useful only if it beats that mixture.

The comparison also tends to omit a cost. A method that gates on closed-book confidence must first let the reader generate a closed-book answer for every query, including those it then sends to retrieval. The latency of this probe has to be charged to the adaptive method.

This paper measures confidence-based allocation against these alternatives. We use a frozen reader that can answer from its own knowledge, from one retrieved passage or from five. An adaptive policy chooses among the three by thresholding the reader's closed-book confidence; a simple policy is a fixed depth or a random mixture of fixed depths. Both kinds are given the same cap on the average number of passages per query, are selected on development questions and are scored on held-out questions. In a stricter version of the comparison, the simple policy may also read two or three passages for every query. We then repeat the comparison at equal latency, with the probe included.

To see what the outcome depends on, we run the same TriviaQA and Natural Questions questions and the same reader with two retrieval sources. In the first, each question comes with its own small pool of candidate passages, which usually contains the answer, and retrieval almost always helps. In the second, passages are retrieved from an index of 3.5 million Wikipedia passages, and the top passages often displace an answer the reader already had. On the Wikipedia index, the same questions are also run with an 8B and a 32B reader.

The answer depends on what a threshold policy is asked to do. It can beat a random mixture in three ways. It can ration a scarce budget, giving the first passage to the queries that need it most. It can avoid harm, skipping retrieval for queries whose closed-book answer a passage would spoil. And it can choose depth, reading five passages only where one is not enough. We find that closed-book confidence supports the first two and barely the third. At a cap of half a passage per query, where some queries must go without retrieval, the adaptive policy beats the best simple policy in all seven primary settings, by \gHalfMin{} to \gHalfMax{} in accuracy. At caps of one passage or more, it keeps a substantial advantage only on Wikipedia, where retrieval spoils many answers the reader knew, and there in two of the four settings; the largest gain is \gTqaWikiETwo, for the 8B reader on TriviaQA. With passage pools it gains almost nothing, and it falls behind once the simple policy may read two or three passages for every query. The reason lies in the signal: confidence ranks which queries a first passage helps, but it ranks the choice between one passage and five little better than chance.

Charging the probe narrows the result further. In our measurements the closed-book probe takes nearly as long as a one-passage answer, so a gate between no passage and one has to skip retrieval for most queries before it saves any time. At equal latency the rationing gains with passage pools disappear. What remains are Wikipedia settings in which the gate skips retrieval for most queries, and for the 32B reader, how much of the gain remains depends on the answer metric.

Besides these findings, the paper offers an evaluation protocol that can be reused for other adaptive methods: compare with simple policies under the same budget cap, including intermediate fixed depths; select every policy on development data; and charge the probe. The study is limited to short-answer question answering with frozen open-weight readers (Section~\ref{sec:practice}).

\section{Related work}\label{sec:related}

\paragraph{Deciding whether to retrieve from the reader's uncertainty.}
Much of the work on adaptive retrieval decides, query by query, whether retrieval is needed, and uses the reader's own uncertainty as the evidence. Active retrieval methods trigger retrieval during generation when token-level uncertainty is high \cite{jiang2023flare,su2024dragin}. Self-RAG fine-tunes the reader to emit reflection tokens that request retrieval and critique passages \cite{asai2024selfrag}, and ADAPT-LLM trains a reader to emit a special token when it does not know the answer \cite{labruna2024when}. Other methods probe the reader's hidden states \cite{baek2024probingrag} or estimate uncertainty from them \cite{yao2024seakr}. Training-free gates threshold uncertainty statistics computed over the first tokens of a short closed-book draft \cite{wang2026targ}. A comparison of 35 adaptive-retrieval and uncertainty-estimation methods found that simple uncertainty estimates often match complex adaptive pipelines in answer quality while being more efficient \cite{moskvoretskii2025adaptive}. Building on that finding, we use one of the simplest estimates, the log-probability of the reader's closed-book answer, and repeat the signal analysis with the prefix statistics of the training-free gates.

\paragraph{Deciding with a dedicated gate.}
A second line builds a separate component whose only task is the retrieval decision. Adaptive-RAG trains a query-complexity classifier that routes queries to no retrieval, single-step retrieval or iterative retrieval \cite{jeong2024adaptiverag}. SKR elicits the model's self-knowledge to decide whether to retrieve \cite{wang2023skr}. RAGate builds a binary gate for conversational turns from human judgements of whether a turn needs retrieved knowledge \cite{wang2025ragate}. Related work outside retrieval also learns, per instance, how much context or computation to spend: adaptive in-context learning predicts how many demonstrations to place in the prompt \cite{chandra2025onesize}, and model routing learns when a cheaper generator suffices \cite{ong2025routellm}. We do not evaluate these gates, but the oracle analysis of Section~\ref{sec:room} applies to them: whatever signal a router uses, it cannot gain more over the best simple policy than an oracle with the same actions and budget on the same questions.

\paragraph{Deciding how much to retrieve.}
Fewer methods choose the number of passages, and those that do mostly rely on evidence from the retriever. Adaptive-$k$ selects the cut-off from the distribution of query--passage similarity scores in a single pass and reports large token savings at matched accuracy \cite{taguchi2025adaptivek}. DynamicRAG trains a reranker that chooses how many documents to keep, using the quality of the reader's output as feedback \cite{sun2025dynamicrag}. A neighbouring line predicts the benefit of a retrieved set: retrieval performance prediction estimates how much a fixed set of retrieved documents improves on answering without context \cite{tian2026predicting}, and query performance predictors have been used to analyse agentic retrieval-augmented generation \cite{tian2025qppagentic}. In two of our pool settings, retriever scores rank the benefit of reading more passages where the reader's confidence does not (Section~\ref{sec:signal}).

\paragraph{When more context hurts.}
More retrieved text sometimes makes an answer worse. Readers use information at the beginning and end of a long context better than information in the middle \cite{liu2024lost}. The accuracy of long-context readers can rise and then fall as passages are added, with retrieved hard negatives as a main cause \cite{jin2024longcontext}, and the type, position and number of passages in the prompt all affect answer quality \cite{cuconasu2024power}. Imperfect retrieval can introduce misleading information that conflicts with what the reader knows \cite{wang2024astute}, and augmenting every conversational turn has been reported to increase generation uncertainty \cite{wang2025ragate}. Errors can also be separated by whether the retrieved context was sufficient to answer \cite{joren2025sufficient}. We measure this effect directly as the \emph{harm rate}: the share of questions that the reader answers correctly without retrieval but incorrectly with it. Its counterpart, the \emph{rescue rate}, is the share answered incorrectly without retrieval but correctly with it. Section~\ref{sec:room} relates the two rates to what a gate can gain.

\paragraph{Confidence and calibration.}
A confidence score can be judged by how well it ranks examples or by how well its values match observed accuracy, and the two can diverge: neural confidence scores are often miscalibrated even when they rank well \cite{guo2017calibration}. Language models' own probabilities carry information about whether an answer is correct \cite{kadavath2022know}, and semantic uncertainty improves on token likelihood for free-form generation \cite{kuhn2023semantic}. For threshold policies only the ranking matters. A calibrator that preserves the order of the scores cannot change which queries fall above a threshold tuned on development data, and we confirm that it changes none of our decisions (Section~\ref{sec:calibration}).

\section{Evaluation design}\label{sec:protocol}

\subsection{Actions and policies}\label{sec:policies}

For a query $q$, the system either answers closed-book, which we write as $k{=}0$, or reads the top $k$ retrieved passages. $Y_k(q)\in\{0,1\}$ records whether the reader's answer with $k$ passages is correct. The adaptive policies choose among $k\in\{0,1,5\}$. Depths of two and three passages appear only in simple policies, as additional baselines.

A simple policy makes no per-query decision. The plainest kind is a \emph{fixed} policy, which reads the same number of passages for every query. A \emph{budget-matched mixture} reads $k_1$ passages for a randomly chosen share of the queries and $k_2$ for the rest, and the share sets its average number of passages. Its expected accuracy is the corresponding weighted average of the two fixed accuracies.

The \emph{best simple policy} at budget $B$ is the most accurate fixed policy or two-depth mixture whose average number of passages does not exceed $B$. Mixtures of more than two depths need not be considered, because under a single constraint on the average depth some mixture of at most two depths is always among the most accurate. In the main comparison it draws on the same depths $\{0,1,5\}$ as the adaptive policies. In the stricter comparison it may also use depths of two and three passages.

An adaptive policy first runs the reader closed-book and takes a confidence score $s(q)$ from that answer. It then sends the most confident queries to the cheapest action. A two-level \emph{gate} $\{a,b\}$, with $a<b$ and a threshold $t$, reads $a$ passages if $s(q)\ge t$ and $b$ passages otherwise. We consider three gates. The gates $\{0,1\}$ and $\{0,5\}$ decide whether to retrieve, and the gate $\{1,5\}$ decides how much to read. The three-level policy makes both decisions with two thresholds: it reads no passage if $s(q)\ge t_0$, one passage if $t_1\le s(q)<t_0$, and five otherwise. Whenever a policy reads no passage for a query, the closed-book answer it has already generated is the final answer.

\subsection{Budget}\label{sec:budget}

Policies are compared under a budget $B$, a cap on the average number of passages per query, with $B\in\{0.5,1,2,3\}$. A cap below one cannot be met by retrieving for every query, so at $B{=}0.5$ every policy must leave some queries without retrieval. The cap is enforced on development data: a candidate policy is admissible if its average depth on the development questions does not exceed $B$. The selected policy is then applied unchanged to held-out questions, and we report what it actually reads.

A policy need not spend its whole budget. If more passages do not raise development accuracy, the best simple policy stays below the cap, and the adaptive policy may then read more than it does. Section~\ref{sec:cost} therefore also compares the two at equal latency.

\subsection{Data, readers and retrieval}\label{sec:data}

Table~\ref{tab:settings} lists the seven primary settings. Each uses 2,000 questions drawn from TriviaQA \cite{joshi2017triviaqa}, Natural Questions \cite{kwiatkowski2019nq} or MS MARCO \cite{nguyen2016msmarco}; removing duplicate questions leaves 1,997 for TriviaQA. The readers are Qwen3-8B and Qwen3-32B \cite{qwen3technicalreport}, used without fine-tuning, with greedy decoding, thinking mode disabled and at most 24 new tokens. The closed-book prompt asks for a short answer from the model's own knowledge. The open-book prompt lists the passages and asks for a short answer. No prompt with five or fewer passages reached the reader's input limit of 4,096 tokens in the primary settings; the further settings used a limit of 2,048 tokens.

\begin{table*}[t]
\caption{Primary settings. Columns $k{=}0$ to $k{=}5$ give accuracy with that many passages. Rescue and harm are the rates at which five passages turn a wrong closed-book answer into a correct one and a correct one into a wrong one. NQ: Natural Questions; later tables and figures write 8B and 32B for Qwen3-8B and Qwen3-32B.}
\label{tab:settings}
\centering\small
\resizebox{\linewidth}{!}{\input{tab_settings}}
\end{table*}

The two retrieval settings use the same encoder, BGE-large (bge-large-en-v1.5) \cite{xiao2024cpack}, and differ in what it searches. In the \emph{pool} setting it ranks only the candidate passages that come with each question: the evidence documents for TriviaQA, split into chunks of at most 1,200 characters, the passages released with Dense Passage Retrieval for Natural Questions \cite{karpukhin2020dpr}, and the provided passages for MS MARCO. These pools usually contain the answer. In the \emph{Wikipedia} setting the encoder searches a flat index of 3,539,275 Wikipedia passages, a much harder retrieval task.

Table~\ref{tab:settings} shows how differently the two settings affect the reader. For Qwen3-8B on TriviaQA, five pool passages raise accuracy from \accTqaPoolKZero{} closed-book to \accTqaPoolKFive, whereas five Wikipedia passages raise it only to \accTqaWikiEKFive. In no primary setting does accuracy fall beyond five passages: with ten it is at least as high as with five (\nTruncTen{} ten-passage prompts were truncated).

Some pools are small. \nShortTqaPool{} of the 1,997 TriviaQA questions have fewer than five candidate passages, so the action $k{=}5$ reads \meanPassFiveTqaPool{} passages on average there; \nShortMsmPool{} questions are affected in MS MARCO and none in Natural Questions. The budget $B$ counts the nominal depth, whereas the costs in Section~\ref{sec:cost} use the passages and tokens actually read.

Fifteen further settings, with 300 to 600 questions each, widen the coverage. They add HotpotQA \cite{yang2018hotpotqa}, the readers Qwen3-1.7B \cite{qwen3technicalreport}, Qwen3.5-9B \cite{qwen35modelcard} and Llama-3.1-8B \cite{llama3herd}, a weaker BGE-small retriever (bge-small-en-v1.5) and a shared-corpus variant of the pools. They are reported in Appendix~\ref{app:breadth}.

\subsection{Confidence signal and answer metric}\label{sec:signalmetric}

The confidence score $s(q)$ is the length-normalised log-probability of the closed-book answer. The signal analysis of Section~\ref{sec:signal} also uses the entropy of the opening tokens of the closed-book answer and the margin between their two most probable tokens, as in training-free gates \cite{wang2026targ}, the length of the question, and a logistic combination of all of these signals.

An answer counts as correct if, after lower-casing and removing punctuation and articles, it equals or contains one of the gold answer strings. \emph{Accuracy} refers to this criterion throughout. Because the criterion accepts a long answer that happens to contain a gold string, we repeat the main comparison with exact match and token-level F1, selecting policies to maximise each metric in turn (Section~\ref{sec:metric}).

\subsection{Selection and evaluation}\label{sec:selection}

Every policy, simple or adaptive, is selected and evaluated by the same procedure. We split the questions into five folds and treat each fold in turn as held-out data, with the other four as development data. Fold membership is a hash of the question text and of a repeat index. Duplicated questions therefore always fall in the same fold, and each repeat index gives a new split. We average the held-out results over twenty splits.

On the development folds we choose, for each kind of policy and each budget, the candidate with the highest development accuracy among those whose development depth does not exceed $B$. Ties go to the candidate that reads fewer passages. For threshold policies, the candidate thresholds are 41 quantiles of the development scores. The chosen policy is applied once to the held-out fold. \emph{The adaptive policy} at budget $B$ is whichever of the three gates and the three-level policy has the highest development accuracy.

Policies are compared by the difference in their held-out accuracy on the same questions. Confidence intervals come from an outer bootstrap that resamples questions, keeps copies of a question in the same fold, and repeats the entire selection procedure for each resample (200 replicates; Appendix~\ref{app:stats}). The intervals therefore include the variability of choosing thresholds as well as that of sampling questions. We call a difference significant when its 95\% interval excludes zero.

\section{When does adaptive allocation help?}\label{sec:pays}

\begin{figure*}[t]
\centering
\includegraphics[width=\linewidth]{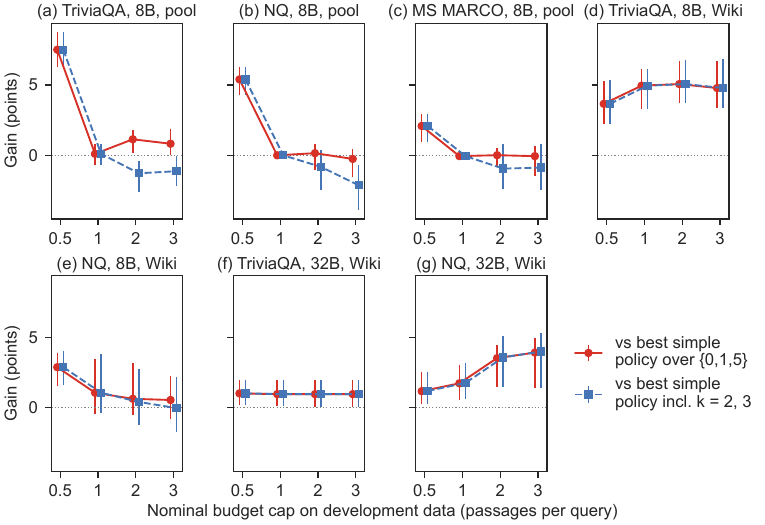}
\caption{Held-out accuracy of the adaptive policy minus the best simple policy, in accuracy points (100 times the differences given in the text and tables), against the budget cap, for the seven primary settings (Wiki: Wikipedia index). Circles, solid lines: simple policies over $\{0,1,5\}$, the same actions as the adaptive policy. Squares, dashed lines: simple policies that may also read two or three passages for every query. Bars are 95\% outer-bootstrap intervals that include threshold selection}
\label{fig:gains}
\end{figure*}

Figure~\ref{fig:gains} and Table~\ref{tab:gains} report, for each setting and budget, the held-out accuracy of the adaptive policy minus that of the best simple policy selected under the same budget cap. The last column of the table is discussed in Section~\ref{sec:levels}.

\begin{sidewaystable}
\caption{Held-out accuracy of the adaptive policy minus the best simple policy (fixed depth or budget-matched mixture over $\{0,1,5\}$) at budget caps of $B$ passages per query, with 95\% outer-bootstrap intervals. Bold: interval excludes zero. The last column compares the three-level policy with the best two-level gate.}
\label{tab:gains}
\centering\footnotesize
\resizebox{\linewidth}{!}{\input{tab_gains}}
\end{sidewaystable}

\subsection{Below one passage per query}\label{sec:below}

At a cap of half a passage per query, the adaptive policy beats the best simple policy in all seven settings, and every gain is significant. The gains range from \gHalfMin{} to \gHalfMax. They are largest on the TriviaQA and Natural Questions pools (\gTqaPoolHalf{} and \gNqPoolHalf), the two settings in which passages rescue the most answers.

In the pools, the best simple policy at this cap retrieves one passage for a random half of the queries. The adaptive policy also retrieves one passage for half of the queries, but for the half in whose closed-book answers the reader is least confident. We refer to this as the \emph{rationing} gain. It arises even where retrieval rarely harms an answer.

\subsection{One passage per query or more: passage pools}\label{sec:pools}

With passage pools, the gain all but disappears once the cap reaches one passage per query. The adaptive policy is ahead by at most \maxPoolGain{} (TriviaQA at two passages), and for Natural Questions and MS MARCO no gain at a cap of one to three passages is significant.

The simple policy becomes harder to beat when it may use intermediate depths. If it may also read two or three passages, the adaptive policy over $\{0,1,5\}$ falls behind it. The difference is \gAllkTqaPoolTwo{} on the TriviaQA pools at a cap of two passages (interval \gAllkTqaPoolTwoLo{} to \gAllkTqaPoolTwoHi) and \gAllkNqPoolThree{} on Natural Questions at a cap of three (Figure~\ref{fig:gains}, squares; Table~\ref{tab:gainsallk} in Appendix~\ref{app:breadth}). A fixed intermediate depth is thus a stronger baseline than the always-five system with which adaptive methods are usually compared.

\subsection{One passage per query or more: Wikipedia}\label{sec:wiki}

On Wikipedia the adaptive policy keeps its advantage at higher budgets, most clearly for Qwen3-8B on TriviaQA. There it gains \gTqaWikiEOne, \gTqaWikiETwo{} and \gTqaWikiEThree{} at caps of one, two and three passages, and allowing the simple policy to read two or three passages changes nothing (Figure~\ref{fig:gains}d). The adaptive policy retrieves only for the queries with the least confident closed-book answers, skipping retrieval for \skipTqaWikiETwo{} of the queries at a cap of two, and this is more accurate than any fixed depth. A gain that does not depend on a tight budget can arise only where retrieval harms some answers (Section~\ref{sec:room}), and we refer to it as the \emph{harm-avoidance} gain.

The other three Wikipedia settings gain less, and not always significantly. Qwen3-32B on Natural Questions gains \gNqWikiTTOne{} at a cap of one passage and \gNqWikiTTThree{} at a cap of three, again whether or not the simple policy may use intermediate depths. Qwen3-32B on TriviaQA gains a small but significant \gTqaWikiTTTwo{} at every cap. For this reader the best simple policy is never to retrieve (\accTqaWikiTTKZero{} closed-book against \accTqaWikiTTKFive{} with five passages), and the adaptive policy improves on it by retrieving for a small minority of queries, reading \passAdaptTqaWikiTTTwo{} passages per query at a cap of two. Qwen3-8B on Natural Questions gains \gNqWikiETwo{} at a cap of two passages, which is not significant.

A shared cap does not mean equal spending. For Qwen3-32B on Natural Questions at a cap of two, the adaptive policy reads \passAdaptNqWikiTTTwo{} passages per query and the simple policies selected on development data only \passSimpleNqWikiTTTwo{} on average, because for this reader five passages are on average no more accurate than none (Table~\ref{tab:settings}). The gain at this cap therefore comes at a higher cost, which the comparison at equal latency in Section~\ref{sec:cost} takes into account.

\subsection{Sensitivity to the answer metric}\label{sec:metric}

Table~\ref{tab:emf1} repeats the comparison at a cap of two passages per query with exact match and F1, with policies selected and scored by each metric. The main positive case does not depend on the metric. For Qwen3-8B on TriviaQA Wikipedia the adaptive policy gains \emGainTqaWikiE{} in exact match and \fOneGainTqaWikiE{} in F1, about the same as in accuracy. The pool results, small gains or none, are also unchanged.

\begin{table*}[t]
\caption{Exact match (EM) and token F1. Left: F1 of fixed policies. Right: held-out gain of the adaptive policy over the best simple policy at two passages per query when policies are selected and scored with each metric (accuracy: outer-bootstrap interval, bold where it excludes zero; EM and F1: query-bootstrap interval conditional on the selected policies, not marked).}
\label{tab:emf1}
\centering\scriptsize
\resizebox{\linewidth}{!}{\input{tab_emf1}}
\end{table*}

The 32B reader behaves differently. For Qwen3-32B on Natural Questions Wikipedia, the gain shrinks from \gNqWikiTTTwo{} in accuracy to \emGainNqWikiTT{} in exact match and \fOneGainNqWikiTT{} in F1. Qwen3-32B answers at greater length when it reads passages, which accuracy accepts as long as the answer contains a gold string, whereas exact match rejects it and F1 penalises it. Under F1, retrieval is clearly harmful on average for this reader (\fOneNqWikiTTKZero{} closed-book against \fOneNqWikiTTKFive{} with five passages), and the gate adds almost nothing to never retrieving. This gain thus rests on the lenient accuracy criterion; the gain of the 8B reader on TriviaQA does not.

\section{Why the gains differ}\label{sec:why}

An adaptive policy can gain only if two conditions hold: some allocation under the same cap must be better than the simple policy's, and the confidence signal must be able to find it. We begin with the signal.

\subsection{What a gate has to rank}\label{sec:identity}

Consider a gate and a mixture that spend the same budget. The gate sends a set $S$ of queries to action $a$ and the rest to action $b$. The mixture sends the same share of queries to $a$, but picks them at random. Let $D(q)=Y_a(q)-Y_b(q)$ be the per-query gain of $a$ over $b$: it is $+1$ if only $a$ gives a correct answer, $-1$ if only $b$ does, and $0$ otherwise. The gate's accuracy then exceeds the mixture's expected accuracy by
\begin{equation}
  \mathrm{Acc}(\text{gate}) - \mathrm{Acc}(\text{mixture}) \;=\; \mathrm{Cov}\big(\mathbf{1}[q\in S],\, D(q)\big),
  \label{eq:cov}
\end{equation}
with the covariance taken over queries. In words, the gate beats the mixture exactly to the extent that the queries it sends to $a$ gain more from $a$, on average, than queries in general. Whatever signal a gate uses, the signal must rank the per-query gain of the step that the gate controls.

We therefore ask how well confidence ranks the gain of each retrieval step: the first from no passage to one, the second from one passage to five. We measure each step in the direction of adding passages: its gain is correctness after the step minus correctness before it, so that $+1$ is a rescue and $-1$ a harm. For each step we take the queries whose correctness the step changes and compute the area under the receiver operating characteristic curve (AUROC) with which low confidence separates the rescues from the harms; 0.5 means no ordering. The AUROC is a diagnostic and not a full account of a gate's value, since it leaves aside the queries whose correctness does not change and summarises a monotone ranking.

\subsection{Confidence ranks the first passage, not the depth}\label{sec:signal}

\begin{figure}[t]
\centering
\includegraphics[width=\linewidth]{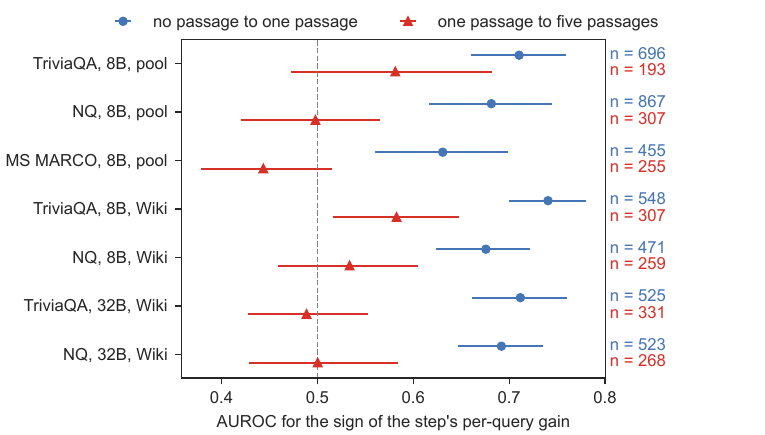}
\caption{AUROC with which low closed-book confidence orders the queries whose answer a retrieval step changes, for the step from no passage to one (circles) and from one passage to five (triangles), with 95\% bootstrap intervals. The dashed line marks no ranking. Right margin: number of queries whose answer the step changes, of about 2,000 per setting (Wiki: Wikipedia index)}
\label{fig:signal}
\end{figure}

Confidence ranks the first step in every setting (Figure~\ref{fig:signal}, circles). The AUROC for the step from no passage to one lies between \aucStepOneLo{} and \aucStepOneHi, and every interval excludes 0.5. When the reader is confident in its closed-book answer, the answer is usually right, so a passage has little to rescue and something to spoil.

The second step is different (Figure~\ref{fig:signal}, triangles). For the step from one passage to five, the same signal reaches an AUROC of only \aucStepTwoLo{} to \aucStepTwoHi, and most intervals include 0.5 (the step changes correctness for \nChangeTwoLo{} to \nChangeTwoHi{} queries per setting). Other signals available before retrieval do no better: the entropy of the opening tokens behaves like the log-probability, and a logistic model that combines all pre-retrieval signals reaches \aucComboTwoLo{} to \aucComboTwoHi{} (Table~\ref{tab:signal}).

Among the primary settings, only Qwen3-8B on TriviaQA Wikipedia shows some ranking of the second step (\aucSeqTqaWikiETwo{}, interval \aucSeqTqaWikiETwoLo--\aucSeqTqaWikiETwoHi). Among the fifteen further settings, Qwen3-32B and Llama-3.1-8B on the TriviaQA pools do (\aucTqaTTPoolSmall{} and \aucLlamaPool, intervals \aucTqaTTPoolSmallLo--\aucTqaTTPoolSmallHi{} and \aucLlamaPoolLo--\aucLlamaPoolHi), but the 32B effect did not reappear on Wikipedia, and for Qwen3-8B on Natural Questions Wikipedia the ranking is reversed (\aucNqWikiSmall, interval \aucNqWikiSmallLo--\aucNqWikiSmallHi).

Where the reader's confidence fails, the retriever's scores sometimes help. In the Natural Questions and MS MARCO pools, the similarity scores of the retrieved passages rank the gain of the second step with an AUROC of \aucRetNqPool{} (interval \aucRetNqPoolLo--\aucRetNqPoolHi) and \aucRetMsmPool{} (\aucRetMsmPoolLo--\aucRetMsmPoolHi), where closed-book confidence reaches \aucSeqNqPool{} and \aucSeqMsmPool. The effect is absent on the TriviaQA pools and on Wikipedia. Where it appears, it suggests using the reader's confidence to decide whether to retrieve and the retriever's scores to decide how much to read, in line with retriever-side methods for choosing $k$ \cite{taguchi2025adaptivek}. Retriever scores exist only after retrieval, so they cannot inform the decision to skip it.

\subsection{Where there is something to gain}\label{sec:room}

A signal can only find gains that exist. The simplest case has two actions, no passage or $k$ passages, and a budget large enough to retrieve for every query.

\begin{proposition}[Unbudgeted headroom]\label{prop:headroom}
Let $D=Y_k-Y_0$, and let $\rho^{+}=\Pr(D=1)$ and $\rho^{-}=\Pr(D=-1)$ be the rates at which reading $k$ passages rescues and harms an answer. The best per-query choice between $0$ and $k$ passages, without a budget, exceeds the better of the two fixed policies by exactly $\min(\rho^{+},\rho^{-})$.
\end{proposition}

\begin{proof}
The per-query oracle is correct whenever either action is correct, so its accuracy is $\Pr(Y_0{=}1 \vee Y_k{=}1)=\bar Y_0+\rho^{+}=\bar Y_k+\rho^{-}$. Subtracting $\max(\bar Y_0,\bar Y_k)$ leaves $\min(\rho^{+},\rho^{-})$. For a real-valued outcome such as F1, the same argument gives $\min(\mathbb{E}[D^{+}],\mathbb{E}[D^{-}])$ with $D^{\pm}=\max(\pm D,0)$.
\end{proof}

The proposition says that choosing per query whether to retrieve can add much only when retrieval both rescues and harms many answers. If harms are rare, always retrieving is nearly as good as the oracle, and if rescues are rare, never retrieving is. This separates our two retrieval settings. In the pools, five passages harm only \harmNqPool{} to \harmTqaPool{} of the answers. On Wikipedia, five passages rescue \rescTqaWikiE{} and harm \harmTqaWikiE{} of the answers of Qwen3-8B on TriviaQA, and rescue \rescNqWikiTT{} and harm \harmNqWikiTT{} of those of Qwen3-32B on Natural Questions. Harm avoidance thus has something to work with on Wikipedia and very little in the pools.

This unbudgeted headroom covers harm avoidance only. When the budget is too small to retrieve for every query, rationing is a second source of gain, and it is not limited by the unbudgeted headroom: at half a passage per query the realised gain exceeds the headroom for five passages in \nPropExceedHalf{} of \nOracleSettings{} settings (Table~\ref{tab:halfall}).

To measure what is available from both sources, at any budget and with all three actions, we compare the best simple policy with three oracles. Each knows the outcome of every action for every query and respects the budget. The \emph{full} oracle allocates freely over $\{0,1,5\}$. The \emph{skip-only} oracle may use only $\{0,1\}$ or only $\{0,5\}$, whichever gains more, so it decides whom to skip. The \emph{depth-only} oracle may use only $\{1,5\}$. It reads at least one passage for every query and decides only how deep to read, so it is defined for caps of at least one passage. For an action set $\mathcal{A}$, the value of the oracle is
\begin{equation}
  \begin{aligned}
  V^{\mathrm{oracle}}_{\mathcal{A}}(B) \;=\; &\max_{x}\ \frac{1}{n}\sum_{q}\sum_{a\in\mathcal{A}} x_{qa}\,Y_a(q)\\
  &\text{s.t.}\quad \sum_{a} x_{qa}=1,\ \ x_{qa}\ge 0,\ \ \frac{1}{n}\sum_{q}\sum_{a} a\,x_{qa}\le B,
  \end{aligned}
  \label{eq:oracle}
\end{equation}
a linear programme in which $x_{qa}$ is the probability of taking action $a$ for query $q$. We call the value of an oracle minus the accuracy of the best simple policy over $\{0,1,5\}$ at the same budget the \emph{room} for adaptive allocation: the full room, the skip-only room and the depth-only room. All are computed on all questions of a setting.

The restricted rooms do not add up to the full room. A query can count in both, and a restricted room can even be negative: for Qwen3-32B on TriviaQA Wikipedia, the depth-only oracle loses to never retrieving. All rooms are diagnostics computed with every outcome known. They do not strictly bound the realised gain, which is measured on held-out folds against simple policies selected on development data, yet that gain exceeded the full room in \nOracleExceed{} of \nOracleCases{} combinations of setting and budget.

\begin{figure*}[t]
\centering
\includegraphics[width=\linewidth]{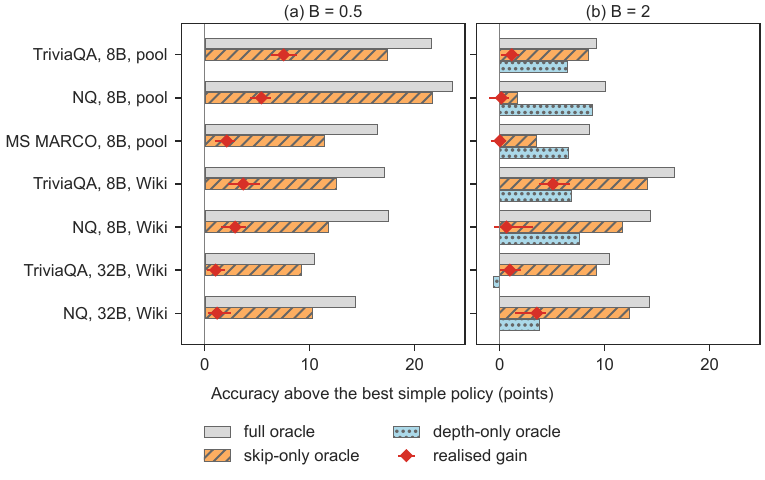}
\caption{Room for adaptive allocation over the best simple policy at caps of (a) 0.5 and (b) 2 passages per query, in accuracy points, from restricted-action oracle comparisons: the full oracle over $\{0,1,5\}$ (plain bars), an oracle restricted to the decision whom to skip (hatched; $\{0,1\}$ or $\{0,5\}$) and one restricted to the decision how deep to read (dotted; $\{1,5\}$, defined only for caps of at least one passage). The restricted comparisons are not additive. Diamonds: the realised held-out gain of the adaptive policy with its 95\% outer-bootstrap interval. Wiki: Wikipedia index. Numbers in Table~\ref{tab:room}}
\label{fig:room}
\end{figure*}

At half a passage per query the room is large in every setting, between \roomAllTqaWikiTTHalf{} and \roomAllNqPoolHalf{} (Figure~\ref{fig:room}a), and the skip-only oracle reaches most of it. With less than one passage per query, any simple policy leaves many queries without a passage that would have rescued them, in the pools as much as on Wikipedia. This is the room behind the rationing gain, and since confidence ranks the first step, the adaptive policy collects part of it in every setting: between \shareNqWikiTTHalf{} and \shareTqaPoolHalf{} of the full room.

At two passages per query the settings separate (Figure~\ref{fig:room}b). In the Natural Questions and MS MARCO pools, the skip-only oracle gains little over the best simple policy (\roomSkipNqPoolTwo{} and \roomSkipMsmPoolTwo), whereas the depth-only oracle gains \roomDepthNqPoolTwo{} and \roomDepthMsmPoolTwo. What could still be gained there lies in the choice between one passage and five. That is the step closed-book confidence does not rank, and the adaptive policy gains nothing. On Wikipedia the skip-only room stays large (\roomSkipTqaWikiTTTwo{} to \roomSkipTqaWikiETwo), because retrieval harms many answers the reader knew, and that is a step confidence does rank.

At two passages per query the adaptive policy realises \shareTqaWikiETwo{} of the full room for Qwen3-8B on TriviaQA Wikipedia and \shareNqWikiTTTwo{} for Qwen3-32B on Natural Questions Wikipedia, but only \shareNqWikiETwo{} for Qwen3-8B on Natural Questions Wikipedia, where the room is just as large. One difference is that retrieval helps this last reader on average, raising its accuracy from \accNqWikiEKZero{} to \accNqWikiEKFive{} with five passages. Skipping a query at random would then forgo an expected benefit, which the ranking has to outweigh. Room is necessary for a gain without guaranteeing one.

Across all \nOracleSettings{} settings, primary and further, the gain at two passages per query follows the skip-only room (Spearman \rhoSkipRoomTwo) more closely than the full room (\rhoAllRoomTwo) or the depth-only room (\rhoDepthRoomTwo). The settings share questions and readers, so these correlations are descriptive.

\subsection{Two levels or three}\label{sec:levels}

If confidence does not rank the depth step, a third level should add little, and that is what we find. In none of the seven primary settings does the three-level policy differ significantly from the best two-level gate (Table~\ref{tab:gains}, last column). The largest difference in either direction, at any cap from one to three passages, is \maxAbsThreeVsTwo. Among the fifteen further settings there is one significant difference, and it goes against the three-level policy: Qwen3-1.7B on the TriviaQA pools at three passages per query (Table~\ref{tab:halfall}).

A three-level policy needs its signal to support both decisions at once, whom to skip and how deep to read, whereas a two-level gate can rely on the one decision the signal ranks. Three levels could pay off with a signal that also ranks the depth step, for which retriever scores are a candidate in the Natural Questions and MS MARCO pools (Section~\ref{sec:signal}).

\subsection{Calibration}\label{sec:calibration}

Calibrating the confidence score does not change any of these results. A logistic calibrator maps the score to a probability that the closed-book answer is correct, and its expected calibration error is far lower than that of the raw score rescaled to the unit interval (Table~\ref{tab:calibration}). But the calibrator preserves the order of the scores, and the thresholds are chosen among development quantiles, so the same queries fall above the selected threshold on either scale. The held-out decisions were identical in all \nCalCheck{} combinations of setting and signal that we checked. Calibration matters for rules that compare a probability with a fixed cost ratio.

\section{Do the gains survive the probe's cost?}\label{sec:cost}

The comparisons so far count passages and leave out the closed-book probe that an adaptive policy runs for every query. We measured latency on an NVIDIA H100 PCIe GPU at batch size one, with Hugging Face Transformers in bfloat16, for the probe and for open-book generation with one and five passages, fitted timing models to these runs and applied them to the token counts of every query in every setting (Appendix~\ref{app:cost}).

The probe costs nearly as much as a one-passage answer (Table~\ref{tab:cost}). Averaged over each setting's queries, the timing models put it at \probeTqaE{} ms on TriviaQA and \probeNqE{} ms on Natural Questions for Qwen3-8B, close to the \probeMeasTqaE{} and \probeMeasNqE{} ms of the queries that were actually timed. For Qwen3-32B the models give \probeTqaTT{} and \probeNqTT{} ms, and the timed queries averaged \probeMeasTqaTT{} and \probeMeasNqTT{} ms. The reason is that generation time is dominated by the number of generated tokens, at about \msPerGenTokE{} ms per token for Qwen3-8B and \msPerGenTokTT{} ms for Qwen3-32B, and much less by the length of the prompt.

\begin{table*}[t]
\caption{Mean latency per query estimated from timing models fitted to measured runs (batch size one, H100 PCIe), averaged over each setting's queries; open-book latencies include retrieval. Break-even skip rate: the share of queries a gate $\{0,k\}$ must skip to be faster than always reading $k$ passages. Gain at equal latency: held-out accuracy of the adaptive policy minus the best mixture of fixed policies over $\{0,1,5\}$ with the same expected latency (a hindsight envelope that favours the simple policy). MS MARCO was not timed; its costs are predicted from its token counts.}
\label{tab:cost}
\centering\scriptsize
\resizebox{\linewidth}{!}{\input{tab_cost}}
\end{table*}

We cost a policy query by query, along the route the policy chose. Every query of an adaptive policy pays for the probe. A query sent to retrieval also pays for retrieval and for an open-book pass predicted from its own token counts; a query not sent to retrieval reuses the probe's answer. Retrieval is modelled as a separate service with a GPU-resident index, at \retrMsOne{} ms for one passage and \retrMsFive{} ms for five. Simple policies pay no probe.

\subsection{How many queries a gate must skip}\label{sec:breakeven}

Compare a gate between zero and $k$ passages with always reading $k$ passages. The gate pays the probe on every query and saves the cost of retrieving and reading on the queries it skips. If the skipped queries cost about as much to retrieve and read as the average query, the gate is faster only if the share of queries it skips exceeds the ratio of the probe's latency to the latency of retrieving and reading $k$ passages. For a gate $\{0,1\}$ this break-even skip rate is \breakEvenEightLo{} to \breakEvenEightHi{} for Qwen3-8B and \breakEvenTTLo{} to \breakEvenTTHi{} for Qwen3-32B. For a gate $\{0,5\}$ with Qwen3-8B it is \breakEvenFiveLo{} to \breakEvenFiveHi{} (Table~\ref{tab:cost}). In the pools the selected policies skip far fewer queries than that. At a cap of two passages the adaptive policy skips retrieval for \skipTqaPoolTwo{} of the queries on TriviaQA and for none on Natural Questions and MS MARCO.

\subsection{Gains at equal latency}\label{sec:equallatency}

\begin{figure*}[t]
\centering
\includegraphics[width=\linewidth]{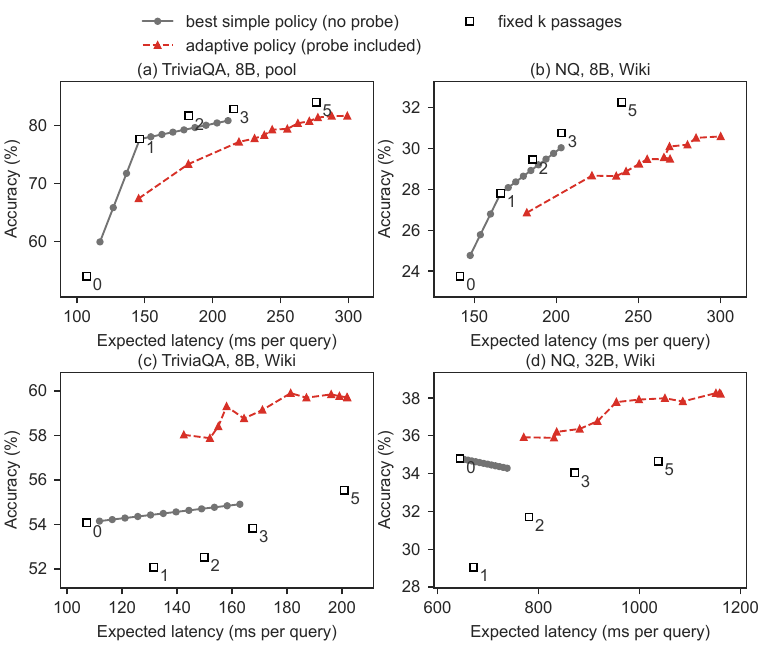}
\caption{Held-out accuracy against expected latency per query for the best simple policy over $\{0,1,5\}$ (circles; no probe) and the adaptive policy (triangles; probe included), both selected on development data, at budget caps from 0.25 to 3 passages per query in steps of 0.25. Open squares: fixed policies, labelled with their number of passages. Wiki: Wikipedia index}
\label{fig:cost}
\end{figure*}

Table~\ref{tab:cost} also compares the adaptive policy with simple policies at equal expected latency. The reference in the table is the best fixed policy or mixture over $\{0,1,5\}$ whose expected latency does not exceed that of the adaptive policy, chosen with the accuracies of the fixed policies on all questions. Figure~\ref{fig:cost} instead plots simple policies selected on development data; against those, the differences change by at most \latSweepMaxDiff, always in the adaptive policy's favour.

From one passage per query upward, the outcome at equal latency follows the split of Section~\ref{sec:pays}, only more sharply. In the pools the adaptive policy now loses, with differences between \latPoolLo{} and \latPoolHi{} at caps of one and two passages, and it also loses for Qwen3-8B on Natural Questions Wikipedia. In the other three Wikipedia settings it keeps its gain: \latTqaWikiETwo{} for Qwen3-8B on TriviaQA, \latNqWikiTTTwo{} for Qwen3-32B on Natural Questions and \latTqaWikiTTTwo{} for Qwen3-32B on TriviaQA at a cap of two passages. In these three settings the gate skips retrieval for most queries (\skipTqaWikiETwo, \skipNqWikiTTTwo{} and \skipTqaWikiTTTwo{} of them), so the probe replaces most open-book passes instead of being added to them.

Below one passage per query, adaptive allocation gained in every setting under the same passage cap. At equal latency the gain survives only in the same three Wikipedia settings. In the pools and for Qwen3-8B on Natural Questions Wikipedia the probe turns it into a loss: at a cap of half a passage the difference is \latTqaPoolHalf{} on the TriviaQA pools and \latNqWikiEHalf{} for Qwen3-8B on Natural Questions Wikipedia. With passage pools, the rationing gain is therefore a saving in passages and not in time: a system that can afford the probe's latency does better to spend it on a passage for every query. In Figure~\ref{fig:cost} the adaptive policy lies below the simple policies at every latency on the TriviaQA pools, and above every fixed depth for Qwen3-8B on TriviaQA Wikipedia.

These costs hold for single-query decoding with Hugging Face Transformers and a separate retrieval service, and the MS MARCO costs are predicted from token counts, not timed. Serving systems that batch requests or reuse the probe's prefix would lower the probe's relative cost; the break-even condition can be evaluated with the costs of the deployed system.

\section{Practical implications and limitations}\label{sec:practice}

\subsection{Checking a new deployment}\label{sec:checks}

The results suggest three questions to ask before building a confidence-based allocator. Is the budget below one passage per query? If so, allocation by confidence gained in all seven primary settings, but the gain is counted in passages and, with pools, does not survive the probe's latency. Does retrieval change many answers in both directions? The rescue and harm rates of a labelled sample and the skip-only room show whether a gate has anything to find at larger budgets; where harms are rare, a fixed depth of two or three passages is hard to beat. Is the probe cheap relative to an open-book pass on the deployed system? If not, a gate has to skip retrieval for most queries to pay for itself.

We tested how far a small labelled pilot can make this choice automatically. A deployment usually asks which policy is cheapest without giving up accuracy, so the pilot rule minimises latency within an accuracy tolerance. It labels 400 queries, runs the reader closed-book and with one and five passages, and considers the fixed policies with zero, one and five passages and the gates $\{0,1\}$ and $\{0,5\}$, each gate at every threshold among the quantiles of the pilot's scores. It chooses the candidate with the lowest mean latency whose pilot accuracy is within $\delta=0.01$ of the best fixed policy. A conservative variant requires, for every fixed policy, that the paired accuracy difference between the candidate and that policy stays above $-\delta$ at its one-sided lower bound (the mean minus 1.645 standard errors); if no candidate passes, it falls back to the fixed policy that is most accurate on the pilot.

How reliable the pilot is depends on how clearly one candidate stands out (Table~\ref{tab:pilot}). We compare the choice of each pilot with the choice the rule makes when given all queries of the setting. In the TriviaQA and Natural Questions pools, where the full-data choice is five passages, conservative pilots agree with it in \agreeConsNqPool{} to \agreeConsTqaPool{} of 200 draws. The hardest case for the tolerance is Qwen3-8B on TriviaQA Wikipedia. The full-data choice there is the gate $\{0,1\}$, but several candidates lie close to the tolerance; never retrieving, for instance, is only slightly less accurate than five passages (\accTqaWikiEKZero{} against \accTqaWikiEKFive). Standard pilots agree with the full-data choice in only \agreeStdTqaWikiE{} of draws and conservative pilots in \agreeConsTqaWikiE. The chosen policy loses more than $\delta$ to the best fixed policy on the remaining queries in \violStdTqaWikiE{} of draws under the standard rule and \violConsTqaWikiE{} under the conservative one. With Qwen3-32B on Natural Questions Wikipedia, conservative pilots seldom match the full-data choice (\agreeConsNqWikiTT{} of draws), but they pick more accurate and slower policies rather than violate the tolerance. Because the conservative rule screens many thresholds on the same pilot, its bound does not guarantee a 5\% violation rate, and a pilot of this size is a diagnostic that narrows the choice.

\subsection{How much development data threshold selection needs}\label{sec:devsize}

\begin{figure}[t]
\centering
\includegraphics[width=\linewidth]{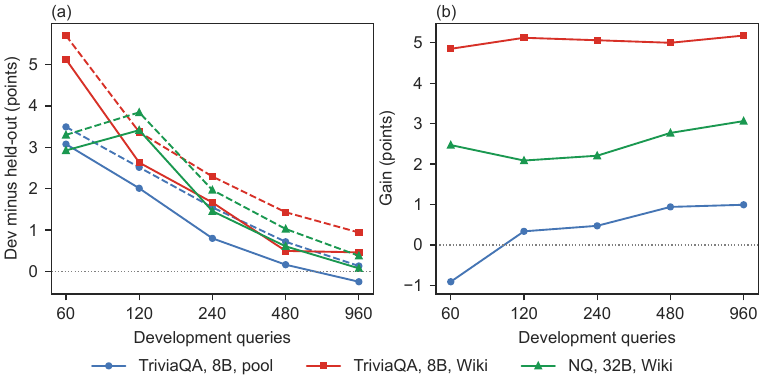}
\caption{(a) Development minus held-out accuracy of the selected policy at two passages per query, for two-level (solid) and three-level (dashed) policies. (b) Held-out gain of the adaptive policy over the best simple policy. Both against the number of development queries used for selection, for three representative settings (Wiki: Wikipedia index)}
\label{fig:stability}
\end{figure}

Thresholds chosen on a small development set overfit, and two thresholds overfit more than one. We measure this as the optimism of selection, the development accuracy of the chosen policy minus its held-out accuracy (Figure~\ref{fig:stability}; Table~\ref{tab:optimism}). For Qwen3-8B on TriviaQA Wikipedia, optimism falls from \gapTwoLevelSmall{} for two-level gates and \gapThreeLevelSmall{} for the three-level policy with 120 development queries to \gapTwoLevelLarge{} and \gapThreeLevelLarge{} with 960. Two-level gates are less optimistic at every size and, as Section~\ref{sec:levels} showed, no less accurate, which makes them the better default.

In the pools, an adaptive policy selected on a small development set loses to the best simple policy: on TriviaQA with 60 queries, and on Natural Questions and MS MARCO with fewer than 960. In the four Wikipedia settings it gains at every size (Table~\ref{tab:devgain}).

\subsection{Limitations}\label{sec:limitations}

The study covers short-answer factoid question answering with frozen open-weight readers. Long-form generation, conversational settings and multi-step retrieval may behave differently. The 2,000-question Wikipedia settings use one reader family, Qwen3, and for its larger model the conclusions depend on the answer metric. Each setting also uses a single retriever. A different retriever would change the rescue and harm rates and with them the room, so these quantities should be measured on a labelled sample from the deployment.

Several measurements have limits of their own. The latencies hold for the serving setup of Section~\ref{sec:cost}. The budget counts nominal depth (Section~\ref{sec:data}): at a cap of two, for example, the best simple policy reads \passActSimpleTqaPoolTwo{} passages per query on the TriviaQA pools and the adaptive policy \passActAdaptTqaPoolTwo. The fifteen further settings come from earlier runs whose pools were built slightly differently from those of the 2,000-question runs, so questions from the two sets of runs are never combined. Finally, the MS MARCO labels are noisier than the others: a manual audit of a sample found an error rate of 12.5\%, so small differences on MS MARCO are less reliable.

\section{Conclusion}\label{sec:conclusion}

We asked when allocating retrieved passages by the reader's closed-book confidence pays off against a fixed depth or a random mixture of fixed depths under the same budget cap. It can pay off in two situations. The first is a budget below one passage per query: confidence ranks which queries a first passage helps, so it rations a scarce budget better than chance in all seven primary settings. This gain is a saving in passages, and with passage pools it does not survive the latency of the probe. The second is retrieval that changes many answers in both directions, as happens when retrieval is weak relative to what the reader already knows. A gate that skips retrieval for most queries can then beat every fixed depth, and in three of our four Wikipedia settings it still does so after its probe is paid for, although for the larger reader how much of the gain remains depends on the answer metric. Elsewhere the best simple policy, often a fixed depth of two or three passages, is as good or better.

Closed-book confidence ranks whether to retrieve much better than how much to read. Where an oracle finds most of the remaining room in the choice of depth, as in the Natural Questions and MS MARCO pools at two passages per query, confidence cannot reach it, and a third level added no significant gain to a two-level gate. The rescue and harm rates of retrieval, the room an oracle leaves for each decision, and the cost of the probe relative to an open-book pass can all be measured on a labelled sample. They show where a gate can gain; whether it does still has to be checked on held-out queries.

\section*{Code and data availability}
The code, per-query tables, timing records and results are available at \url{https://github.com/dongzhe1/know-before-you-fetch}.

\section*{Acknowledgements}
We thank the anonymous reviewers of an earlier version of this work for their detailed comments.

\appendix

\section{Additional settings}\label{app:breadth}

\begin{figure}[h]
\centering
\includegraphics[width=0.78\linewidth]{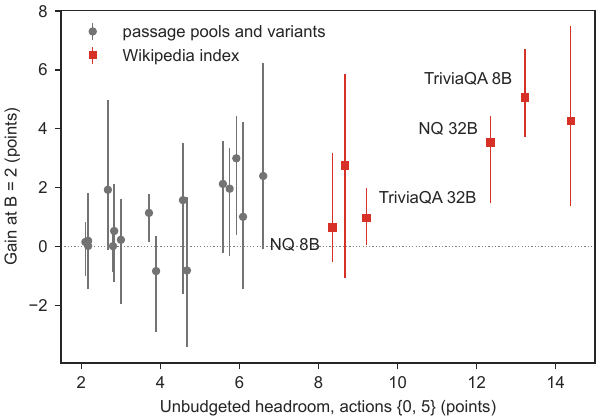}
\caption{Held-out gain of the adaptive policy over the best simple policy at two passages per query (95\% outer-bootstrap intervals) against the unbudgeted headroom $\min(\rho^{+},\rho^{-})$ of Proposition~\ref{prop:headroom} for the actions $\{0,5\}$, for all \nOracleSettings{} settings. The headroom is a descriptive measure of how complementary the two actions are, not a bound for this budgeted gain (Spearman \rhoHeadroom; the settings share questions and readers)}
\label{fig:headroom}
\end{figure}

Table~\ref{tab:breadth} reports the fifteen further settings with 300 to 600 questions. The pattern matches the primary settings: gains at two passages per query are mostly small and not significant with passage pools, and largest for TriviaQA on Wikipedia, where the room for the skip decision is large. Table~\ref{tab:gainsallk} gives the comparison with simple policies that may also read two or three passages, and Figure~\ref{fig:headroom} relates the realised gains to the unbudgeted headroom of Proposition~\ref{prop:headroom}.

\begin{table*}[h]
\caption{The fifteen further settings, from earlier runs with 300 to 600 questions (591 for TriviaQA after removing duplicates) whose pools were built slightly differently; the two Wikipedia rows are therefore not the primary Wikipedia settings. Unbudgeted headroom: $\min(\rho^{+},\rho^{-})$ for five passages (Proposition~\ref{prop:headroom}). Room at $B{=}2$: budgeted oracle minus best simple policy over $\{0,1,5\}$, with the full action set (full) and restricted to the skip decision (skip; Section~\ref{sec:room}). Last column: held-out gain of the adaptive policy over the best simple policy at two passages per query with its 95\% outer-bootstrap interval.}
\label{tab:breadth}
\centering\scriptsize
\resizebox{\linewidth}{!}{\input{tab_breadth}}
\end{table*}

\begin{table*}[h]
\caption{Adaptive policy over $\{0,1,5\}$ minus the best simple policy that may also read two or three passages (95\% outer-bootstrap intervals).}
\label{tab:gainsallk}
\centering\scriptsize
\resizebox{0.85\linewidth}{!}{\input{tab_gains_allk}}
\end{table*}

\section{Statistical procedure}\label{app:stats}

Fold membership is a hash of the normalised question and the repeat index, so a question lies in the same fold in every setting. Point estimates average the held-out gain over twenty repeats of five-fold selection. The outer bootstrap draws questions with replacement, assigns copies of a question to the same fold, re-runs the whole selection with four repeats and records the mean gain; intervals are the 2.5th and 97.5th percentiles of 200 replicates. A query-level bootstrap that holds the selected policies fixed gives intervals about a fifth narrower; we use it only for the exact-match and F1 comparison (Table~\ref{tab:emf1}), where it is labelled as conditional on the selected policies. The budgeted oracle of Equation~\eqref{eq:oracle} is solved exactly with the HiGHS linear-programming solver.

\section{Cost model}\label{app:cost}

We timed 300 queries per reader and dataset on the Wikipedia setting of TriviaQA and Natural Questions. Each query was timed for the closed-book probe, including the log-probabilities the gate needs, and for open-book generation with one and five passages, in random order after a warm-up. Closed-book and open-book latency were each modelled as a linear function of prompt and generated tokens, fitted to these timings, and applied to the measured token counts of every query in every setting. Retrieval latency, query encoding plus exact search on a GPU-resident index of the Wikipedia passages, comes from a separate batch-one benchmark. MS MARCO was not timed, and its costs come from the same fitted models applied to its token counts.

\section{Supplementary tables}\label{app:tables}

This appendix collects seven further tables: the ranking of four signals for both retrieval steps (Table~\ref{tab:signal}; Section~\ref{sec:signal}), the room for adaptive allocation and its restricted-action comparisons (Table~\ref{tab:room}; Section~\ref{sec:room}), the evaluation of the pilot rule (Table~\ref{tab:pilot}; Section~\ref{sec:checks}), the optimism of threshold selection (Table~\ref{tab:optimism}; Section~\ref{sec:devsize}), the calibration of the confidence score (Table~\ref{tab:calibration}; Section~\ref{sec:calibration}), the held-out gain by development-set size (Table~\ref{tab:devgain}; Section~\ref{sec:devsize}), and all \nOracleSettings{} settings at a cap of half a passage per query (Table~\ref{tab:halfall}; Sections~\ref{sec:room} and~\ref{sec:levels}).

\begin{table*}[htbp]
\caption{How well each signal orders the sign of the per-query gain of a retrieval step (AUROC with 95\% bootstrap interval; bold where the interval excludes 0.5). Seq.\ log-prob: length-normalised log-probability of the closed-book answer; Prefix entropy: mean token entropy of the closed-book answer; All pre-retrieval: logistic combination of the pre-retrieval signals; Retriever scores: similarity scores of the retrieved passages, available only after retrieval.}
\label{tab:signal}
\centering\scriptsize
\resizebox{\linewidth}{!}{\input{tab_signal}}
\end{table*}

\begin{table*}[htbp]
\caption{Room for adaptive allocation at caps of 0.5 and 2 passages per query. Full: budgeted oracle over $\{0,1,5\}$ minus the best simple policy; Skip and Depth: the same with the oracle restricted to the skip decision or to the depth decision (separate comparisons, not additive); Gain: realised held-out gain of the adaptive policy; Share: Gain divided by Full.}
\label{tab:room}
\centering\scriptsize
\resizebox{\linewidth}{!}{\input{tab_room}}
\end{table*}

\begin{table*}[htbp]
\caption{Pilot rule (400 labelled queries, $\delta{=}0.01$, 200 draws). Full-data choice: the rule applied to all remaining queries; std.\ and cons.: standard and conservative rule. Agreement: share of pilots that make the same choice. Violation rate: share of pilots whose choice loses more than $\delta$ to the best fixed policy on the remaining queries. Accuracy loss and latency saved (ms per query) relative to the best fixed policy, averaged over pilots, for the conservative rule; costs use mean latencies per action.}
\label{tab:pilot}
\centering\scriptsize
\resizebox{\linewidth}{!}{\input{tab_pilot}}
\end{table*}

\begin{table*}[htbp]
\caption{Optimism of threshold selection: development minus held-out accuracy of the selected policy at two passages per query, by development-set size.}
\label{tab:optimism}
\centering\scriptsize
\resizebox{0.8\linewidth}{!}{\input{tab_optimism}}
\end{table*}

\begin{table*}[htbp]
\caption{Calibration of closed-book log-probability as a predictor of a correct closed-book answer (held-out, five folds): raw score after min--max scaling versus logistic calibration, measured by the expected calibration error (ECE) and the Brier score. Identical decisions: held-out policy decisions coincide for every budget and policy; the check covers all \nCalCheck{} combinations of setting and signal, of which the seven primary settings are shown.}
\label{tab:calibration}
\centering\scriptsize
\resizebox{0.85\linewidth}{!}{\input{tab_calibration}}
\end{table*}

\begin{table*}[htbp]
\caption{Held-out gain of the adaptive policy over the best simple policy at two passages per query, by the number of development queries used for selection (mean over draws).}
\label{tab:devgain}
\centering\small
\input{tab_devgain}
\end{table*}

\begin{table*}[htbp]
\caption{All \nOracleSettings{} settings at a cap of half a passage per query: unbudgeted headroom $\min(\rho^{+},\rho^{-})$ for five passages (Proposition~\ref{prop:headroom}) and held-out gain of the adaptive policy over the best simple policy; last column: three-level policy minus the best two-level gate at three passages per query. Intervals: 95\% outer bootstrap, bold where they exclude zero. Upper block: primary settings; lower block: further settings.}
\label{tab:halfall}
\centering\scriptsize
\resizebox{\linewidth}{!}{\input{tab_half_all}}
\end{table*}

\FloatBarrier

\bibliographystyle{plainnat}
\bibliography{refs}

\end{document}

%% file: numbers.tex
\newcommand{\accNqWikiEKFive}{0.323}

\newcommand{\accNqWikiEKZero}{0.237}

\newcommand{\accTqaPoolKFive}{0.841}

\newcommand{\accTqaPoolKZero}{0.541}
\newcommand{\accTqaWikiEKFive}{0.555}

\newcommand{\accTqaWikiEKZero}{0.541}
\newcommand{\accTqaWikiTTKFive}{0.616}

\newcommand{\accTqaWikiTTKZero}{0.682}

\newcommand{\agreeConsNqPool}{0.98}

\newcommand{\agreeConsNqWikiTT}{0.08}
\newcommand{\agreeConsTqaPool}{1.00}
\newcommand{\agreeConsTqaWikiE}{0.34}

\newcommand{\agreeStdTqaWikiE}{0.17}

\newcommand{\aucComboTwoHi}{0.60}
\newcommand{\aucComboTwoLo}{0.41}
\newcommand{\aucLlamaPool}{0.70}
\newcommand{\aucLlamaPoolHi}{0.87}
\newcommand{\aucLlamaPoolLo}{0.53}

\newcommand{\aucNqWikiSmall}{0.28}
\newcommand{\aucNqWikiSmallHi}{0.42}
\newcommand{\aucNqWikiSmallLo}{0.14}
\newcommand{\aucRetMsmPool}{0.61}
\newcommand{\aucRetMsmPoolHi}{0.68}
\newcommand{\aucRetMsmPoolLo}{0.54}
\newcommand{\aucRetNqPool}{0.69}
\newcommand{\aucRetNqPoolHi}{0.75}
\newcommand{\aucRetNqPoolLo}{0.63}

\newcommand{\aucSeqMsmPool}{0.44}

\newcommand{\aucSeqNqPool}{0.50}

\newcommand{\aucSeqTqaWikiETwo}{0.58}
\newcommand{\aucSeqTqaWikiETwoHi}{0.65}
\newcommand{\aucSeqTqaWikiETwoLo}{0.52}
\newcommand{\aucStepOneHi}{0.74}
\newcommand{\aucStepOneLo}{0.63}
\newcommand{\aucStepTwoHi}{0.58}
\newcommand{\aucStepTwoLo}{0.44}
\newcommand{\aucTqaTTPoolSmall}{0.77}
\newcommand{\aucTqaTTPoolSmallHi}{0.92}
\newcommand{\aucTqaTTPoolSmallLo}{0.60}
\newcommand{\breakEvenEightHi}{0.94}
\newcommand{\breakEvenEightLo}{0.73}
\newcommand{\breakEvenFiveHi}{0.67}
\newcommand{\breakEvenFiveLo}{0.39}
\newcommand{\breakEvenTTHi}{0.96}
\newcommand{\breakEvenTTLo}{0.79}

\newcommand{\emGainNqWikiTT}{+0.006}
\newcommand{\emGainTqaWikiE}{+0.053}
\newcommand{\fOneGainNqWikiTT}{+0.004}
\newcommand{\fOneGainTqaWikiE}{+0.049}
\newcommand{\fOneNqWikiTTKFive}{0.215}
\newcommand{\fOneNqWikiTTKZero}{0.264}

\newcommand{\gAllkNqPoolThree}{$-$0.021}

\newcommand{\gAllkTqaPoolTwo}{$-$0.013}
\newcommand{\gAllkTqaPoolTwoHi}{$-$0.004}
\newcommand{\gAllkTqaPoolTwoLo}{$-$0.026}
\newcommand{\gHalfMax}{0.075}
\newcommand{\gHalfMin}{0.010}

\newcommand{\gNqPoolHalf}{+0.054}

\newcommand{\gNqWikiETwo}{+0.006}

\newcommand{\gNqWikiTTOne}{+0.017}

\newcommand{\gNqWikiTTThree}{+0.039}

\newcommand{\gNqWikiTTTwo}{+0.035}

\newcommand{\gTqaPoolHalf}{+0.075}

\newcommand{\gTqaWikiEOne}{+0.049}

\newcommand{\gTqaWikiEThree}{+0.048}

\newcommand{\gTqaWikiETwo}{+0.051}

\newcommand{\gTqaWikiTTTwo}{+0.010}

\newcommand{\gapThreeLevelLarge}{0.009}
\newcommand{\gapThreeLevelSmall}{0.034}
\newcommand{\gapTwoLevelLarge}{0.005}
\newcommand{\gapTwoLevelSmall}{0.026}
\newcommand{\harmNqPool}{0.021}

\newcommand{\harmNqWikiTT}{0.125}
\newcommand{\harmTqaPool}{0.037}
\newcommand{\harmTqaWikiE}{0.132}

\newcommand{\latNqWikiEHalf}{$-$0.025}

\newcommand{\latNqWikiTTTwo}{+0.032}
\newcommand{\latPoolHi}{$-$0.004}
\newcommand{\latPoolLo}{$-$0.045}
\newcommand{\latSweepMaxDiff}{0.008}
\newcommand{\latTqaPoolHalf}{$-$0.061}

\newcommand{\latTqaWikiETwo}{+0.044}

\newcommand{\latTqaWikiTTTwo}{+0.010}
\newcommand{\maxAbsThreeVsTwo}{0.007}
\newcommand{\maxPoolGain}{0.011}

\newcommand{\meanPassFiveTqaPool}{4.35}
\newcommand{\msPerGenTokE}{27}
\newcommand{\msPerGenTokTT}{49}
\newcommand{\nCalCheck}{50}

\newcommand{\nChangeTwoHi}{331}
\newcommand{\nChangeTwoLo}{193}

\newcommand{\nOracleCases}{88}
\newcommand{\nOracleExceed}{0}
\newcommand{\nOracleSettings}{22}
\newcommand{\nPropExceedHalf}{11}
\newcommand{\nShortMsmPool}{53}
\newcommand{\nShortTqaPool}{516}

\newcommand{\nTruncTen}{2}
\newcommand{\passActAdaptTqaPoolTwo}{1.72}
\newcommand{\passActSimpleTqaPoolTwo}{1.84}
\newcommand{\passAdaptNqWikiTTTwo}{1.93}

\newcommand{\passAdaptTqaWikiTTTwo}{0.53}
\newcommand{\passSimpleNqWikiTTTwo}{0.79}

\newcommand{\probeMeasNqE}{134}
\newcommand{\probeMeasNqTT}{602}
\newcommand{\probeMeasTqaE}{111}
\newcommand{\probeMeasTqaTT}{389}
\newcommand{\probeNqE}{141}
\newcommand{\probeNqTT}{645}
\newcommand{\probeTqaE}{107}
\newcommand{\probeTqaTT}{394}

\newcommand{\rescNqWikiTT}{0.123}

\newcommand{\rescTqaWikiE}{0.147}

\newcommand{\retrMsFive}{27}
\newcommand{\retrMsOne}{18}

\newcommand{\rhoAllRoomTwo}{0.56}
\newcommand{\rhoDepthRoomTwo}{$-$0.20}

\newcommand{\rhoHeadroom}{0.71}

\newcommand{\rhoSkipRoomTwo}{0.81}

\newcommand{\roomAllNqPoolHalf}{0.236}

\newcommand{\roomAllTqaWikiTTHalf}{0.105}

\newcommand{\roomDepthMsmPoolTwo}{+0.065}
\newcommand{\roomDepthNqPoolTwo}{+0.088}

\newcommand{\roomSkipMsmPoolTwo}{0.034}

\newcommand{\roomSkipNqPoolTwo}{0.017}

\newcommand{\roomSkipTqaWikiETwo}{0.141}

\newcommand{\roomSkipTqaWikiTTTwo}{0.092}

\newcommand{\shareNqWikiETwo}{0.04}
\newcommand{\shareNqWikiTTHalf}{0.08}
\newcommand{\shareNqWikiTTTwo}{0.25}
\newcommand{\shareTqaPoolHalf}{0.35}

\newcommand{\shareTqaWikiETwo}{0.30}

\newcommand{\skipNqWikiTTTwo}{0.60}

\newcommand{\skipTqaPoolTwo}{0.19}

\newcommand{\skipTqaWikiETwo}{0.61}

\newcommand{\skipTqaWikiTTTwo}{0.86}

\newcommand{\violConsTqaWikiE}{0.10}

\newcommand{\violStdTqaWikiE}{0.56}

%% file: tab_settings.tex
\begin{tabular}{lllrrrrrrrr}
\toprule
Dataset & Reader & Retrieval & $n$ & $k{=}0$ & $k{=}1$ & $k{=}2$ & $k{=}3$ & $k{=}5$ & Rescue & Harm \\
\midrule
TriviaQA & Qwen3-8B & pool & 1997 & 0.541 & 0.777 & 0.817 & 0.829 & 0.841 & 0.337 & 0.037 \\
NQ & Qwen3-8B & pool & 2000 & 0.237 & 0.614 & 0.637 & 0.655 & 0.659 & 0.443 & 0.021 \\
MS MARCO & Qwen3-8B & pool & 2000 & 0.147 & 0.304 & 0.314 & 0.308 & 0.308 & 0.189 & 0.028 \\
TriviaQA & Qwen3-8B & Wikipedia & 1997 & 0.541 & 0.521 & 0.525 & 0.538 & 0.555 & 0.147 & 0.132 \\
NQ & Qwen3-8B & Wikipedia & 2000 & 0.237 & 0.278 & 0.294 & 0.307 & 0.323 & 0.169 & 0.084 \\
TriviaQA & Qwen3-32B & Wikipedia & 1997 & 0.682 & 0.570 & 0.591 & 0.606 & 0.616 & 0.092 & 0.158 \\
NQ & Qwen3-32B & Wikipedia & 2000 & 0.348 & 0.290 & 0.317 & 0.341 & 0.346 & 0.123 & 0.125 \\
\bottomrule
\end{tabular}

%% file: tab_gains.tex
\begin{tabular}{llllllll}
\toprule
 & & & \multicolumn{4}{c}{Adaptive $-$ best simple policy over $\{0,1,5\}$} & Three-level $-$ \\
\cmidrule(lr){4-7}
Dataset & Reader & Retrieval & $B{=}0.5$ & $B{=}1$ & $B{=}2$ & $B{=}3$ & best two-level ($B{=}2$) \\
\midrule
TriviaQA & 8B & pool & \textbf{+0.075} [+0.063, +0.088] & +0.001 [$-$0.007, +0.008] & \textbf{+0.011} [+0.002, +0.018] & +0.008 [0.000, +0.018] & +0.002 [$-$0.006, +0.008] \\
NQ & 8B & pool & \textbf{+0.054} [+0.043, +0.063] & 0.000 [$-$0.001, 0.000] & +0.002 [$-$0.010, +0.008] & $-$0.003 [$-$0.016, +0.005] & 0.000 [$-$0.001, 0.000] \\
MS MARCO & 8B & pool & \textbf{+0.021} [+0.009, +0.029] & $-$0.001 [$-$0.003, +0.001] & 0.000 [$-$0.009, +0.006] & $-$0.001 [$-$0.015, +0.007] & $-$0.001 [$-$0.003, +0.001] \\
TriviaQA & 8B & Wikipedia & \textbf{+0.037} [+0.023, +0.053] & \textbf{+0.049} [+0.033, +0.061] & \textbf{+0.051} [+0.037, +0.067] & \textbf{+0.048} [+0.033, +0.067] & $-$0.004 [$-$0.012, +0.002] \\
NQ & 8B & Wikipedia & \textbf{+0.029} [+0.016, +0.039] & +0.011 [$-$0.005, +0.035] & +0.006 [$-$0.005, +0.032] & +0.006 [$-$0.008, +0.023] & $-$0.005 [$-$0.011, +0.020] \\
TriviaQA & 32B & Wikipedia & \textbf{+0.010} [+0.002, +0.019] & \textbf{+0.010} [+0.001, +0.020] & \textbf{+0.010} [+0.001, +0.020] & \textbf{+0.010} [+0.001, +0.020] & $-$0.001 [$-$0.005, +0.006] \\
NQ & 32B & Wikipedia & \textbf{+0.012} [+0.003, +0.025] & \textbf{+0.017} [+0.006, +0.030] & \textbf{+0.035} [+0.015, +0.044] & \textbf{+0.039} [+0.014, +0.049] & $-$0.001 [$-$0.006, +0.005] \\
\bottomrule
\end{tabular}

%% file: tab_emf1.tex
\begin{tabular}{lllrrrlll}
\toprule
 & & & \multicolumn{3}{c}{F1 of fixed policies} & \multicolumn{3}{c}{Adaptive $-$ best simple ($B{=}2$)} \\
\cmidrule(lr){4-6}\cmidrule(lr){7-9}
Dataset & Reader & Retrieval & $k{=}0$ & $k{=}1$ & $k{=}5$ & Accuracy & EM & F1 \\
\midrule
TriviaQA & 8B & pool & 0.540 & 0.774 & 0.819 & \textbf{+0.011} [+0.002, +0.018] & +0.009 [+0.001, +0.017] & +0.009 [+0.003, +0.016] \\
NQ & 8B & pool & 0.266 & 0.629 & 0.674 & +0.002 [$-$0.010, +0.008] & $-$0.002 [$-$0.009, +0.004] & +0.002 [$-$0.005, +0.010] \\
MS MARCO & 8B & pool & 0.219 & 0.425 & 0.428 & 0.000 [$-$0.009, +0.006] & $-$0.001 [$-$0.003, +0.001] & +0.002 [$-$0.003, +0.007] \\
TriviaQA & 8B & Wikipedia & 0.540 & 0.534 & 0.559 & \textbf{+0.051} [+0.037, +0.067] & +0.053 [+0.041, +0.065] & +0.049 [+0.038, +0.060] \\
NQ & 8B & Wikipedia & 0.266 & 0.314 & 0.353 & +0.006 [$-$0.005, +0.032] & +0.006 [$-$0.003, +0.014] & +0.006 [$-$0.004, +0.016] \\
TriviaQA & 32B & Wikipedia & 0.589 & 0.475 & 0.477 & \textbf{+0.010} [+0.001, +0.020] & +0.004 [$-$0.001, +0.009] & +0.007 [+0.002, +0.012] \\
NQ & 32B & Wikipedia & 0.264 & 0.218 & 0.215 & \textbf{+0.035} [+0.015, +0.044] & +0.006 [+0.002, +0.010] & +0.004 [$-$0.001, +0.009] \\
\bottomrule
\end{tabular}

%% file: tab_cost.tex
\begin{tabular}{lllrrrrrrrr}
\toprule
 & & & \multicolumn{3}{c}{Latency per query (ms)} & \multicolumn{2}{c}{Break-even skip rate} & \multicolumn{3}{c}{Gain at equal latency} \\
\cmidrule(lr){4-6}\cmidrule(lr){7-8}\cmidrule(lr){9-11}
Dataset & Reader & Retrieval & Probe & $k{=}1$ & $k{=}5$ & $\{0,1\}$ & $\{0,5\}$ & $B{=}0.5$ & $B{=}1$ & $B{=}2$ \\
\midrule
TriviaQA & 8B & pool & 107 & 146 & 276 & 0.73 & 0.39 & $-$0.061 & $-$0.040 & $-$0.030 \\
NQ & 8B & pool & 141 & 166 & 259 & 0.85 & 0.55 & $-$0.162 & $-$0.045 & $-$0.033 \\
MS MARCO & 8B & pool & 185 & 196 & 275 & 0.94 & 0.67 & $-$0.062 & $-$0.005 & $-$0.004 \\
TriviaQA & 8B & Wikipedia & 107 & 132 & 201 & 0.81 & 0.53 & +0.031 & +0.044 & +0.044 \\
NQ & 8B & Wikipedia & 141 & 166 & 240 & 0.85 & 0.59 & $-$0.025 & $-$0.034 & $-$0.027 \\
TriviaQA & 32B & Wikipedia & 394 & 502 & 795 & 0.79 & 0.50 & +0.010 & +0.010 & +0.010 \\
NQ & 32B & Wikipedia & 645 & 671 & 1037 & 0.96 & 0.62 & +0.011 & +0.016 & +0.032 \\
\bottomrule
\end{tabular}

%% file: tab_breadth.tex
\begin{tabular}{lllrrrrrrl}
\toprule
 & & & & & & Unbudgeted & \multicolumn{2}{c}{Room at $B{=}2$} & \\
\cmidrule(lr){8-9}
Dataset & Reader & Retrieval & $n$ & $k{=}0$ & $k{=}5$ & headroom & full & skip & Adaptive $-$ best simple ($B{=}2$) \\
\midrule
TriviaQA & Qwen3-8B & pool & 591 & 0.570 & 0.787 & 0.056 & 0.110 & 0.100 & +0.021 [$-$0.002, +0.036] \\
NQ & Qwen3-8B & pool & 600 & 0.247 & 0.685 & 0.028 & 0.116 & 0.023 & +0.005 [$-$0.012, +0.021] \\
MS MARCO & Qwen3-8B & pool & 600 & 0.142 & 0.310 & 0.022 & 0.080 & 0.030 & +0.002 [$-$0.014, +0.018] \\
HotpotQA & Qwen3-8B & pool & 600 & 0.218 & 0.605 & 0.027 & 0.210 & 0.172 & +0.019 [$-$0.001, +0.050] \\
TriviaQA & Qwen3-1.7B & pool & 591 & 0.271 & 0.672 & 0.039 & 0.115 & 0.050 & $-$0.008 [$-$0.029, +0.003] \\
TriviaQA & Qwen3-32B & pool & 591 & 0.692 & 0.819 & 0.061 & 0.093 & 0.091 & +0.010 [$-$0.014, +0.042] \\
NQ & Qwen3-32B & pool & 300 & 0.343 & 0.670 & 0.047 & 0.114 & 0.074 & $-$0.008 [$-$0.034, +0.017] \\
TriviaQA & Qwen3.5-9B & pool & 591 & 0.638 & 0.800 & 0.046 & 0.086 & 0.076 & +0.016 [$-$0.016, +0.035] \\
TriviaQA & Llama-3.1-8B & pool & 591 & 0.716 & 0.794 & 0.066 & 0.101 & 0.093 & +0.024 [$-$0.001, +0.062] \\
TriviaQA & Qwen3-8B & pool, BGE-small & 591 & 0.570 & 0.783 & 0.058 & 0.122 & 0.117 & +0.020 [$-$0.003, +0.034] \\
TriviaQA & Qwen3-8B & shared corpus & 591 & 0.570 & 0.770 & 0.059 & 0.118 & 0.102 & \textbf{+0.030} [+0.004, +0.044] \\
NQ & Qwen3-8B & shared corpus & 600 & 0.247 & 0.653 & 0.030 & 0.113 & 0.031 & +0.002 [$-$0.020, +0.016] \\
MS MARCO & Qwen3-8B & shared corpus & 600 & 0.142 & 0.300 & 0.022 & 0.075 & 0.032 & 0.000 [$-$0.015, +0.013] \\
TriviaQA & Qwen3-8B & Wikipedia & 591 & 0.570 & 0.591 & 0.144 & 0.176 & 0.156 & \textbf{+0.043} [+0.014, +0.075] \\
NQ & Qwen3-8B & Wikipedia & 600 & 0.247 & 0.312 & 0.087 & 0.148 & 0.118 & +0.027 [$-$0.011, +0.059] \\
\bottomrule
\end{tabular}

%% file: tab_gains_allk.tex
\begin{tabular}{llllll}
\toprule
Dataset & Reader & Retrieval & $B{=}1$ & $B{=}2$ & $B{=}3$ \\
\midrule
TriviaQA & 8B & pool & +0.001 [$-$0.007, +0.008] & \textbf{$-$0.013} [$-$0.026, $-$0.004] & \textbf{$-$0.011} [$-$0.022, 0.000] \\
NQ & 8B & pool & 0.000 [$-$0.001, 0.000] & $-$0.008 [$-$0.025, +0.004] & \textbf{$-$0.021} [$-$0.039, $-$0.007] \\
MS MARCO & 8B & pool & $-$0.001 [$-$0.003, +0.001] & $-$0.009 [$-$0.024, +0.008] & $-$0.009 [$-$0.024, +0.008] \\
TriviaQA & 8B & Wikipedia & \textbf{+0.049} [+0.033, +0.061] & \textbf{+0.051} [+0.037, +0.068] & \textbf{+0.048} [+0.034, +0.068] \\
NQ & 8B & Wikipedia & +0.011 [$-$0.004, +0.038] & +0.004 [$-$0.012, +0.027] & 0.000 [$-$0.017, +0.022] \\
TriviaQA & 32B & Wikipedia & \textbf{+0.010} [+0.001, +0.020] & \textbf{+0.010} [+0.001, +0.020] & \textbf{+0.010} [+0.001, +0.020] \\
NQ & 32B & Wikipedia & \textbf{+0.018} [+0.006, +0.032] & \textbf{+0.036} [+0.015, +0.051] & \textbf{+0.040} [+0.014, +0.053] \\
\bottomrule
\end{tabular}

%% file: tab_signal.tex
\begin{tabular}{lllcccc}
\toprule
 & & & \multicolumn{4}{c}{AUROC for the sign of the per-query gain} \\
\cmidrule(lr){4-7}
Dataset & Reader & Retrieval & Seq.\ log-prob & Prefix entropy & All pre-retrieval & Retriever scores \\
\midrule
\multicolumn{7}{l}{\emph{Step 1: closed book $\rightarrow$ one passage}} \\
TriviaQA & 8B & pool & \textbf{0.71} [0.66, 0.76] & \textbf{0.74} [0.69, 0.78] & \textbf{0.73} [0.69, 0.78] & 0.52 [0.46, 0.57] \\
NQ & 8B & pool & \textbf{0.68} [0.62, 0.74] & \textbf{0.69} [0.63, 0.75] & \textbf{0.70} [0.63, 0.76] & \textbf{0.65} [0.57, 0.72] \\
MS MARCO & 8B & pool & \textbf{0.63} [0.56, 0.70] & \textbf{0.65} [0.58, 0.71] & \textbf{0.59} [0.52, 0.66] & \textbf{0.72} [0.66, 0.78] \\
TriviaQA & 8B & Wikipedia & \textbf{0.74} [0.70, 0.78] & \textbf{0.76} [0.71, 0.80] & \textbf{0.71} [0.67, 0.76] & \textbf{0.73} [0.69, 0.77] \\
NQ & 8B & Wikipedia & \textbf{0.68} [0.62, 0.72] & \textbf{0.69} [0.64, 0.74] & \textbf{0.64} [0.59, 0.69] & \textbf{0.76} [0.71, 0.80] \\
TriviaQA & 32B & Wikipedia & \textbf{0.71} [0.66, 0.76] & \textbf{0.72} [0.67, 0.77] & \textbf{0.72} [0.67, 0.77] & \textbf{0.71} [0.65, 0.76] \\
NQ & 32B & Wikipedia & \textbf{0.69} [0.65, 0.74] & \textbf{0.71} [0.66, 0.75] & \textbf{0.65} [0.60, 0.71] & \textbf{0.74} [0.70, 0.79] \\
\midrule
\multicolumn{7}{l}{\emph{Step 2: one passage $\rightarrow$ five passages}} \\
TriviaQA & 8B & pool & 0.58 [0.47, 0.68] & 0.60 [0.50, 0.70] & 0.60 [0.50, 0.69] & 0.47 [0.34, 0.59] \\
NQ & 8B & pool & 0.50 [0.42, 0.57] & 0.50 [0.43, 0.57] & 0.48 [0.41, 0.54] & \textbf{0.69} [0.63, 0.75] \\
MS MARCO & 8B & pool & 0.44 [0.38, 0.51] & 0.45 [0.38, 0.52] & 0.49 [0.41, 0.56] & \textbf{0.61} [0.54, 0.68] \\
TriviaQA & 8B & Wikipedia & \textbf{0.58} [0.52, 0.65] & \textbf{0.58} [0.51, 0.64] & 0.48 [0.42, 0.55] & 0.54 [0.48, 0.61] \\
NQ & 8B & Wikipedia & 0.53 [0.46, 0.60] & 0.54 [0.46, 0.61] & 0.49 [0.41, 0.56] & 0.57 [0.49, 0.64] \\
TriviaQA & 32B & Wikipedia & 0.49 [0.43, 0.55] & 0.48 [0.42, 0.55] & 0.41 [0.34, 0.48] & 0.55 [0.48, 0.62] \\
NQ & 32B & Wikipedia & 0.50 [0.43, 0.58] & 0.52 [0.45, 0.59] & 0.45 [0.38, 0.52] & 0.52 [0.44, 0.59] \\
\bottomrule
\end{tabular}

%% file: tab_room.tex
\begin{tabular}{lllrrrrrrrrr}
\toprule
 & & & \multicolumn{4}{c}{$B{=}0.5$} & \multicolumn{5}{c}{$B{=}2$} \\
\cmidrule(lr){4-7}\cmidrule(lr){8-12}
Dataset & Reader & Retrieval & Full & Skip & Gain & Share & Full & Skip & Depth & Gain & Share \\
\midrule
TriviaQA & 8B & pool & 0.216 & 0.174 & +0.075 & 0.35 & 0.092 & 0.085 & +0.064 & +0.011 & 0.12 \\
NQ & 8B & pool & 0.236 & 0.217 & +0.054 & 0.23 & 0.101 & 0.017 & +0.088 & +0.002 & 0.02 \\
MS MARCO & 8B & pool & 0.164 & 0.114 & +0.021 & 0.13 & 0.085 & 0.034 & +0.065 & 0.000 & 0.00 \\
TriviaQA & 8B & Wikipedia & 0.171 & 0.126 & +0.037 & 0.21 & 0.166 & 0.141 & +0.068 & +0.051 & 0.30 \\
NQ & 8B & Wikipedia & 0.175 & 0.118 & +0.029 & 0.17 & 0.143 & 0.117 & +0.076 & +0.006 & 0.04 \\
TriviaQA & 32B & Wikipedia & 0.105 & 0.092 & +0.010 & 0.10 & 0.105 & 0.092 & $-$0.006 & +0.010 & 0.09 \\
NQ & 32B & Wikipedia & 0.143 & 0.102 & +0.012 & 0.08 & 0.143 & 0.123 & +0.037 & +0.035 & 0.25 \\
\bottomrule
\end{tabular}

%% file: tab_pilot.tex
\begin{tabular}{llllrrrrrr}
\toprule
 & & & & \multicolumn{2}{c}{Agreement} & \multicolumn{2}{c}{Violation rate} & \multicolumn{2}{c}{Conservative rule} \\
\cmidrule(lr){5-6}\cmidrule(lr){7-8}\cmidrule(lr){9-10}
Dataset & Reader & Retrieval & Full-data choice & std. & cons. & std. & cons. & acc.\ loss & ms saved \\
\midrule
TriviaQA & 8B & pool & fixed $k{=}5$ & 0.98 & 1.00 & 0.01 & 0.00 & 0.000 & 0 \\
NQ & 8B & pool & fixed $k{=}5$ & 0.98 & 0.98 & 0.01 & 0.01 & 0.000 & $-$2 \\
MS MARCO & 8B & pool & fixed $k{=}1$ & 0.57 & 0.35 & 0.10 & 0.10 & +0.004 & +19 \\
TriviaQA & 8B & Wikipedia & gate $\{0,1\}$ & 0.17 & 0.34 & 0.56 & 0.10 & $-$0.021 & +58 \\
NQ & 8B & Wikipedia & fixed $k{=}5$ & 0.71 & 0.94 & 0.28 & 0.03 & +0.001 & $-$2 \\
TriviaQA & 32B & Wikipedia & fixed $k{=}0$ & 1.00 & 0.94 & 0.00 & 0.00 & 0.000 & $-$1 \\
NQ & 32B & Wikipedia & fixed $k{=}0$ & 0.72 & 0.08 & 0.03 & 0.07 & $-$0.016 & $-$102 \\
\bottomrule
\end{tabular}

%% file: tab_optimism.tex
\begin{tabular}{lllrrrr}
\toprule
 & & & \multicolumn{2}{c}{$n_{\mathrm{dev}}{=}120$} & \multicolumn{2}{c}{$n_{\mathrm{dev}}{=}960$} \\
\cmidrule(lr){4-5}\cmidrule(lr){6-7}
Dataset & Reader & Retrieval & two-level & three-level & two-level & three-level \\
\midrule
TriviaQA & 8B & pool & +0.020 & +0.025 & $-$0.002 & +0.001 \\
NQ & 8B & pool & +0.014 & +0.019 & +0.002 & +0.002 \\
MS MARCO & 8B & pool & +0.016 & +0.021 & +0.004 & +0.004 \\
TriviaQA & 8B & Wikipedia & +0.026 & +0.034 & +0.005 & +0.009 \\
NQ & 8B & Wikipedia & +0.037 & +0.038 & +0.006 & +0.009 \\
TriviaQA & 32B & Wikipedia & +0.030 & +0.033 & +0.008 & +0.008 \\
NQ & 32B & Wikipedia & +0.034 & +0.038 & +0.001 & +0.004 \\
\bottomrule
\end{tabular}

%% file: tab_calibration.tex
\begin{tabular}{lllrrrrc}
\toprule
 & & & \multicolumn{2}{c}{ECE} & \multicolumn{2}{c}{Brier} & Identical \\
\cmidrule(lr){4-5}\cmidrule(lr){6-7}
Dataset & Reader & Retrieval & raw & calibrated & raw & calibrated & decisions \\
\midrule
TriviaQA & 8B & pool & 0.334 & 0.061 & 0.309 & 0.182 & yes \\
NQ & 8B & pool & 0.646 & 0.031 & 0.586 & 0.165 & yes \\
MS MARCO & 8B & pool & 0.728 & 0.021 & 0.654 & 0.122 & yes \\
TriviaQA & 8B & Wikipedia & 0.334 & 0.061 & 0.309 & 0.182 & yes \\
NQ & 8B & Wikipedia & 0.646 & 0.031 & 0.586 & 0.165 & yes \\
TriviaQA & 32B & Wikipedia & 0.189 & 0.039 & 0.217 & 0.177 & yes \\
NQ & 32B & Wikipedia & 0.536 & 0.046 & 0.497 & 0.205 & yes \\
\bottomrule
\end{tabular}

%% file: tab_devgain.tex
\begin{tabular}{lllrrrrr}
\toprule
 & & & \multicolumn{5}{c}{Development queries} \\
\cmidrule(lr){4-8}
Dataset & Reader & Retrieval & 60 & 120 & 240 & 480 & 960 \\
\midrule
TriviaQA & 8B & pool & $-$0.009 & +0.003 & +0.005 & +0.009 & +0.010 \\
NQ & 8B & pool & $-$0.018 & $-$0.009 & $-$0.006 & $-$0.002 & +0.001 \\
MS MARCO & 8B & pool & $-$0.018 & $-$0.007 & $-$0.003 & $-$0.002 & 0.000 \\
TriviaQA & 8B & Wikipedia & +0.049 & +0.051 & +0.051 & +0.050 & +0.052 \\
NQ & 8B & Wikipedia & +0.012 & +0.012 & +0.009 & +0.009 & +0.010 \\
TriviaQA & 32B & Wikipedia & +0.015 & +0.011 & +0.006 & +0.010 & +0.010 \\
NQ & 32B & Wikipedia & +0.025 & +0.021 & +0.022 & +0.028 & +0.031 \\
\bottomrule
\end{tabular}

%% file: tab_half_all.tex
\begin{tabular}{lllrrll}
\toprule
 & & & & Unbudgeted & Adaptive $-$ best simple & Three-level $-$ best \\
Dataset & Reader & Retrieval & $n$ & headroom & ($B{=}0.5$) & two-level ($B{=}3$) \\
\midrule
TriviaQA & Qwen3-8B & pool & 1997 & 0.037 & \textbf{+0.075} [+0.063, +0.088] & $-$0.004 [$-$0.007, +0.007] \\
NQ & Qwen3-8B & pool & 2000 & 0.021 & \textbf{+0.054} [+0.043, +0.063] & 0.000 [$-$0.001, 0.000] \\
MS MARCO & Qwen3-8B & pool & 2000 & 0.028 & \textbf{+0.021} [+0.009, +0.029] & $-$0.001 [$-$0.003, +0.001] \\
TriviaQA & Qwen3-8B & Wikipedia & 1997 & 0.132 & \textbf{+0.037} [+0.023, +0.053] & $-$0.002 [$-$0.007, +0.002] \\
NQ & Qwen3-8B & Wikipedia & 2000 & 0.084 & \textbf{+0.029} [+0.016, +0.039] & $-$0.006 [$-$0.012, +0.013] \\
TriviaQA & Qwen3-32B & Wikipedia & 1997 & 0.092 & \textbf{+0.010} [+0.002, +0.019] & $-$0.001 [$-$0.005, +0.006] \\
NQ & Qwen3-32B & Wikipedia & 2000 & 0.123 & \textbf{+0.012} [+0.003, +0.025] & $-$0.001 [$-$0.005, +0.002] \\
\midrule
TriviaQA & Qwen3-8B & pool & 591 & 0.056 & \textbf{+0.071} [+0.037, +0.088] & +0.009 [$-$0.010, +0.034] \\
NQ & Qwen3-8B & pool & 600 & 0.028 & \textbf{+0.050} [+0.024, +0.072] & $-$0.002 [$-$0.010, 0.000] \\
MS MARCO & Qwen3-8B & pool & 600 & 0.022 & \textbf{+0.021} [0.000, +0.040] & 0.000 [$-$0.010, 0.000] \\
HotpotQA & Qwen3-8B & pool & 600 & 0.027 & \textbf{+0.033} [+0.014, +0.050] & +0.015 [$-$0.003, +0.040] \\
TriviaQA & Qwen3-1.7B & pool & 591 & 0.039 & \textbf{+0.065} [+0.037, +0.086] & \textbf{$-$0.003} [$-$0.011, $-$0.001] \\
TriviaQA & Qwen3-32B & pool & 591 & 0.061 & \textbf{+0.042} [+0.014, +0.060] & 0.000 [$-$0.009, +0.012] \\
NQ & Qwen3-32B & pool & 300 & 0.047 & \textbf{+0.059} [+0.018, +0.093] & $-$0.005 [$-$0.022, +0.021] \\
TriviaQA & Qwen3.5-9B & pool & 591 & 0.046 & \textbf{+0.069} [+0.045, +0.089] & $-$0.005 [$-$0.014, +0.019] \\
TriviaQA & Llama-3.1-8B & pool & 591 & 0.066 & \textbf{+0.034} [+0.015, +0.061] & $-$0.005 [$-$0.012, +0.005] \\
TriviaQA & Qwen3-8B & pool, BGE-small & 591 & 0.058 & \textbf{+0.064} [+0.035, +0.082] & +0.002 [$-$0.011, +0.028] \\
TriviaQA & Qwen3-8B & shared corpus & 591 & 0.059 & \textbf{+0.066} [+0.032, +0.088] & +0.017 [$-$0.013, +0.039] \\
NQ & Qwen3-8B & shared corpus & 600 & 0.030 & \textbf{+0.049} [+0.026, +0.068] & $-$0.002 [$-$0.010, 0.000] \\
MS MARCO & Qwen3-8B & shared corpus & 600 & 0.022 & \textbf{+0.021} [0.000, +0.034] & 0.000 [$-$0.009, 0.000] \\
TriviaQA & Qwen3-8B & Wikipedia & 591 & 0.144 & +0.016 [$-$0.004, +0.052] & $-$0.002 [$-$0.014, +0.007] \\
NQ & Qwen3-8B & Wikipedia & 600 & 0.087 & \textbf{+0.053} [+0.024, +0.074] & $-$0.001 [$-$0.011, +0.018] \\
\bottomrule
\end{tabular}